\documentclass{amsart}
\usepackage{amscd}
\usepackage{amssymb}
\usepackage{amsmath,url}
\usepackage{graphicx, bm}

\theoremstyle{definition}

\theoremstyle{remark}

\numberwithin{equation}{section}

\begin{document}

\title[Geometry and Dynamics on Nano Scale]{Towards a Nano Geometry?\\
{Geometry and Dynamics on Nano Scale}}
\author{Bernhelm Boo{\ss}--Bavnbek}
\address{Department of Science, Systems, and Models - NSM/IMFUFA, Roskilde
University, Postboks 260, DK-4000 Ros\-kilde, Denmark}
\email{booss@ruc.dk}
\thanks{{\rm Contribution to the Conference on}
{\em Geometry, Analysis and Quantum Field Theory}, {\rm Potsdam,
September 26-30, 2011}}

\date{January 31, 2012.}

\dedicatory{Dedicated to Steven Rosenberg on his 60th birthday.}


\begin{abstract}
This paper applies I.M. Gelfand's distinction between adequate and
non-adequate use of mathematical language in different contexts to
the newly opened window of model-based measurements of intracellular
dynamics. The specifics of geometry and dynamics on the mesoscale of
cell physiology are elaborated - in contrast to the familiar
Newtonian mechanics and the more recent, but by now also rather well
established quantum field theories. Examples are given originating
from the systems biology of insulin secreting pancreatic beta-cells
and the mathematical challenges of an envisioned non-invasive
control of magnetic nanoparticles.
\end{abstract}

\maketitle

\section{The Challenge of Nano Structures}

There are many different geometries around. Do we really need new
kinds of geometries? Why and how?

\subsection{Nanoparticle Based Transducer for Intracellular Structures}
\quad\newline These days, we witness a dramatic progress in various
technologies devoted to capturing intracellular dynamics of highly
differentiated animal cells like the delicate insulin secreting
pancreatic $\beta$-cell with its thousands of free moving insulin
granules, rails of microtubuli, fences of actin filaments, zoos of
organelles, proteins, genes, ion channels, electro-static and
electrodynamic phenomena. A radically new world of geometry and
dynamics is evolving before our eyes. The most decisive
technological advances are in the following domains:
\begin{itemize}
\item Life imaging, for instance confocal multi-beam laser microscopy,
admitting up to 40 frames per second for tracking position and
movement of suitably prepared nanoparticles within the cell and
without overheating the tissue;
\item Magnetic nanoparticle design and coating, admitting electromagnetic
manipulation, docking to selected organelles and tracking their
movement; \item Computer supported collection and administration of
huge databases.
\end{itemize}
This raises the question: Are we on the verge of a need and, then,
of the emergence of radically different new mathematical concepts,
as predicted by the late I.M. Gelfand in 2003 at a conference (see
\cite{Unity}) on {\em The Unity of Mathematics}, held in honor of
his 90th birthday. He said:
\begin{quotation}
... we have a ``perestroika'' in our time. We have computers which
can do everything. We are not obliged to be bound by two operations
- addition and multiplication. We also have a lot of other tools. I
am sure that in 10 to 15 years mathematics will be absolutely
different from what it was before.  (\cite[p.\/xiv]{Gelfand})
\end{quotation}
Then, how reasonable is it to demand and to expect new geometries
only in view of the new time and length scales of cell physiology:
much larger than the scales underlying particle physics, quantum
mechanics, proteomics and genetics with its characteristic operator
analysis, geometric foldings and stochastic processes; and much
smaller than the scales underlying tissue, organ, patient and
population biology and medicine with its characteristic statistics
and bifurcations? Can't we transfer geometric and dynamic concepts
all the way up and down the scales? What should be so special for
geometers with the mesoscale of a few nanometers and a few seconds
and minutes?

\subsection{Gelfand's Dictum} At the mentioned conference, the jubilee surprised
by {\em qualifying} the general praise of mathematics as adequate
language for science. Against the supposed unity and adequacy of
mathematics he insisted on the distinction between adequate and
inadequate use of mathematical concepts, {\em depending on the
context} (see \cite{Gelfand} for the whole talk):
\begin{quotation}
An important side of mathematics is that it is an adequate language
for different areas: physics, engineering, biology. Here, the most
important word is adequate language. We have adequate and
nonadequate languages. I can give you examples of adequate and
nonadequate languages. For example, to use quantum mechanics in
biology is not an adequate language, but to use mathematics in
studying gene sequences is an adequate language.
\end{quotation}

Clearly, on one side, Gelfand played on the common pride of
mathematicians regarding Galilei's famous dictum of \cite[Il
Saggiatore, cap. 6]{Gal}:
\begin{quotation}
La filosofia {\`e} scritta in questo grandissimo libro che
continuamente ci sta aperto innanzi a gli occhi (io dico
l'universo), ma non si pu{\`o} intendere se prima non s'impara a
intender la lingua, e conoscer i caratteri, ne' quali {\`e} scritto.
Egli {\`e} scritto in lingua matematica, e i caratteri son
triangoli, cerchi, ed altre figure geometriche, senza i quali mezi
{\`e} impossibile a intenderne umanamente parola; senza questi {\`e}
un aggirarsi vanamente per un oscuro laberinto.%
\footnote{In English: ``Philosophy is written in that great book
which ever lies before our eyes — I mean the universe — but we
cannot understand it if we do not first learn the language and grasp
the symbols, in which it is written. This book is written in the
mathematical language, and the symbols are triangles, circles and
other geometrical figures, without whose help it is impossible to
comprehend a single word of it; without which one wanders in vain
through a dark labyrinth." \textit{The Assayer} (1623), as
translated by Thomas Salusbury (1661), p. 178, as quoted in
\textit{The Metaphysical Foundations of Modern Science} (2003) by
Edwin Arthur Burtt, p. 75.}%
\end{quotation}

On the other side, Gelfand warned in the given quote against the
misleading playing around with mathematical concepts without due
regard to the characteristic lengths, times, data and problems of a
concrete context. Is there a contradiction?

\subsection{The Common Regard and Disregard of Context} Deep in our
heart, we mathematicians believe in the unity and universality of
mathematics. We are not {\em topologists}, {\em algebraists}, {\em
pde folk} or {\em applied}, as little as a music composer is a {\em
quartet} or a {\em trio composer}, as Gelfand also noted in his
talk. We are mathematicians, and our belief in the unity and
universality of our concepts is based on three solid pillars,
\begin{enumerate}
\item our Emmy Noether and N. Bourbaki belief in the universal meaning
of structures;
\item our semiotic training which assigns to even the most abstract
concepts very concrete, worldly, human, mental images (a process
intensively studied by the American physicist and philosopher
Charles Sanders Peirce); and
\item our accept of the universality of phenomena,
be it the universality of the three conic sections of Apollonius of
Perga in the level curves of all binary quadratic forms in two
variables $Q(x,y)=a_{11}x^2+2a_{12}xy+a_{22}y^2=c$, or the
universality of Ren\'e Thom's seven elementary catastrophes (generic
structures for the bifurcation geometries) in all dynamical systems
subjected to a potential with two or fewer active variables, and
four or fewer active control parameters.
\end{enumerate}
So, on Sundays we are easily seduced to contempt of the context and
into belief of universality.

However, from the history of our subject we know that there are no
{\em great eternal lines} in mathematics. Euclid did not suffice for
Newton's study of planetary motion, and the calculus was created.
Classical analysis did not suffice for Bohr's study of the atom and
operator theory in Hilbert space was created. Functional analysis
did not suffice for the study of elementary particles and spectral
geometry was developed for the sake of quantum field theories. Worst
of all, there is no mathematics around or emerging in physics to
support a Theory of Everything (TOE) merging all four interactions
into one, in spite of the solid mathematical foundations and the
high promises of the Grand Unified Theory (GUT) to replace the
ad-hoc Standard Model of particle physics. On the contrary, looking
through a modern textbook on {\em Quantum Gravity} like
\cite{Quantum Gravity} will support Niels Bohr's view of the
complementarity and - in tendency - the mutually unrelated state of
different areas of our investigation. So, we have to study a subject
with focused glasses, directed to limited segments, full of
surprises. We have grown used to all kinds of confinements, due to
peculiar aspects of the chosen level of physical reality or due to
fashions, F{\"u}hrers, external impact that can devaluate earlier
approaches and demand radically new ideas over night. So, in daily
work we have learned to live without universality.

\section{Typology of Mathematics Use in Cell Physiology}

\begin{figure}
\includegraphics[scale=0.69]{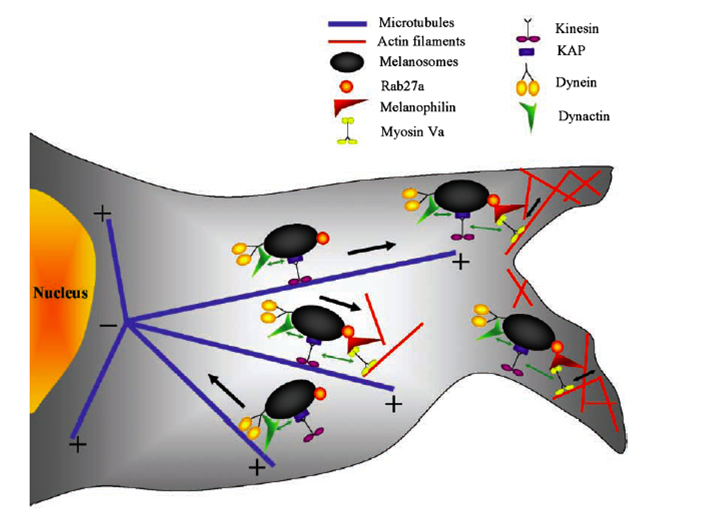}
\caption{Basic model of the transport routes of insulin granules in
pancreatic $\beta$-cells: Before reaching the plasma membrane, the
granules are transferred along microfilaments, courtesy of H.G.
Mannherz} \label{f:motility1}
\end{figure}

\begin{figure}
\includegraphics[scale=0.5]{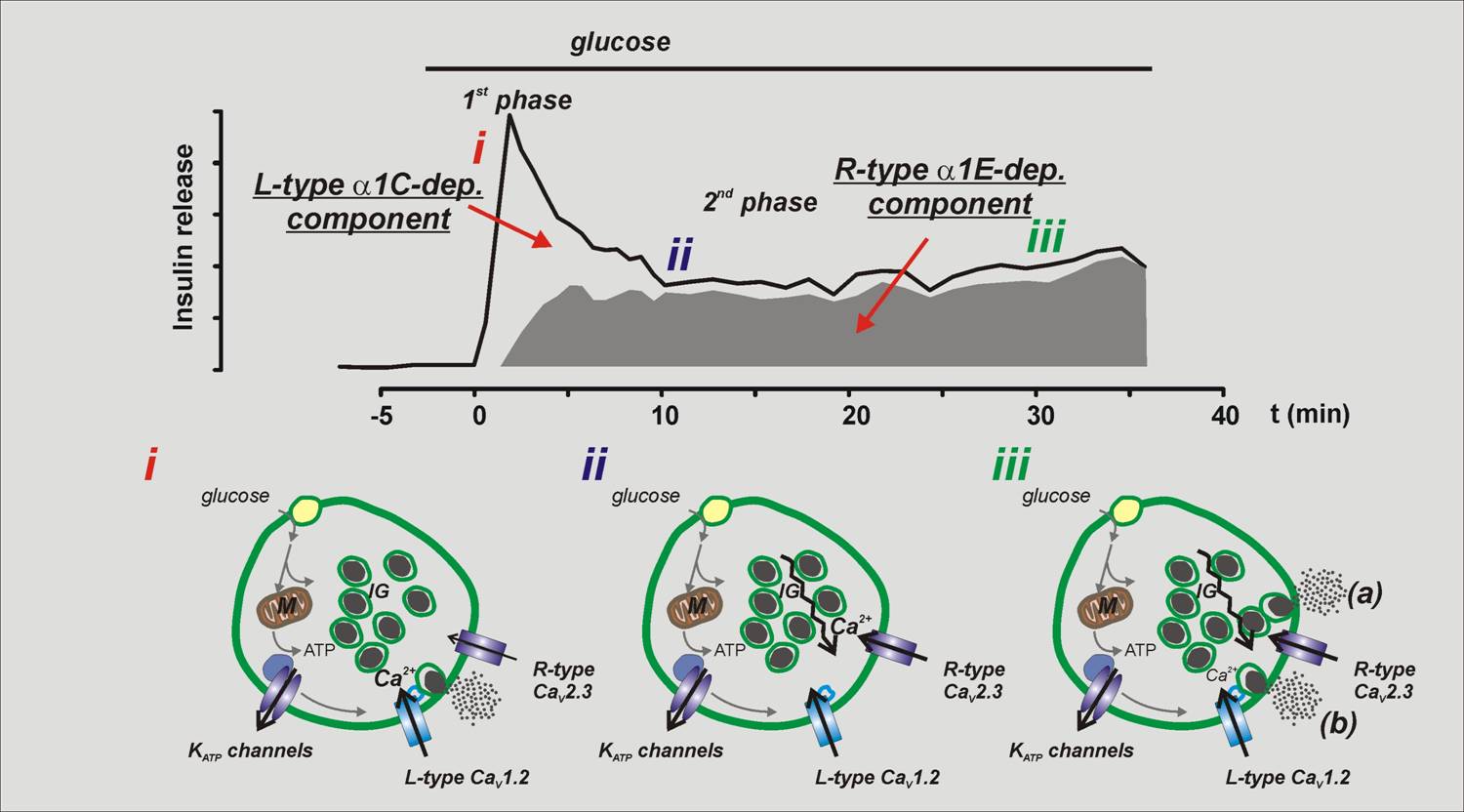}
\caption{Two-phase secretion of insulin with three different
$\beta$-cell modes. The figure shows at the bottom three
$\beta$-cells in three different states. The smaller circles
symbolize insulin vesicles. The graph on top shows the insulin
secretion over time for a single cell. As the graph shows, the
insulin secretion is explosive in the short first phase (mode i). In
the longer second phase (mode iii), the secretion is rather constant
and more evenly distributed. Between the two phases is the waiting
state ii. As depicted in the $\beta$-cells at the bottom of the
figure, the three molecular states are similar to each other.
Consequently, they do not explain the order in the sequence of
phases. It is that order which one now seeks to explain by models of
the underlying geometry and dynamics that involve the interplay
between all processes. After Renstr\"om (2011) in
\cite[p.\/40]{Beta}} \label{f:biphasic}
\end{figure}

For capturing the geometry and the dynamics of insulin secretion of
pancreatic $\beta$-cells ({\em regulated exocytosis}, see Figures
\ref{f:motility1}, \ref{f:biphasic}), it may be helpful to
distinguish the following modeling purposes:

\subsection{Model Based Capturing of Intracellular Dynamics}
With the sudden technology-pushed opening of a window to
intracellular positions, shapes and movements, it seems to me that
the {\em descriptive} role of mathematics will be the most decisive
contribution to the progress of medical biology, i.e., supporting
model-based measurements in the laboratory. To some extent, the
technological progress has given {\em immediate} access to machine
generated cell data in $\beta$-cells like
\begin{itemize}
\item precise measurements
of the quantitative and temporal sequence of glycose-stimulus
secretion-response;
\item precise determination of changes in the electro-static potential
over the plasma membrane and the opening and closing of ion channels
across the plasma membrane upon stimulation;
\item precise observation of positions of
organelles, microfilaments and granules by electron microscopy and
electron tomography under rapid freezing, and vaguely by luminescent
quantum dots and other fluorescent reporters in living cells;
\item identifying proteins, enzymes; and
\item determining genes in DNA sequences.
\end{itemize}
These observations have been around for decades. The drawback with
all of them is their static and local character. No matter how
valuable they are for some purposes, they don't give access to the
intracellular dynamics. So, the true functioning (or dysfunctioning)
of a living $\beta$-cell is not accessible immediately.

Many biomedical quantities cannot be measured directly. That is due
to the subject matter, here the nature of life, partly because most
direct measurements will require some type of fixation, freezing and
killing of the cells, partly due to the small length scale and the
strong interaction between different components of the cell. Just as
in physics since Galileo Galilei's determination of the simple (but
at his time not measurable) free vertical fall law by calculating
``backwards" from the inclined plan, one must also in cell
physiology master the art of model-based experiment design.

Below in Section \ref{ss:viscosity} we shall discuss essential
parameters for the insulin granule motility in $\beta$-cells like
the visco-elasticity of the cytosol or the magnetic field strength
of the pulsating flux of calcium ions between storage organelles
(mitochondria and endoplasmic reticulum). For high precision in the
critical period of granule preparing, docking and bilayer fusion
with the cell membrane, radically new possibilities appear by
tracking the movements of labeled magnetic nanoparticles in
controlled electro-dynamic fields (see below). In this case, solving
mathematical equations from the fields of electro-dynamics and
thermo-elasticity becomes mandatory for the design of the
experiments and the interpretation of the data. In popular terms,
one may speak of a {\em mathematical microscope} (a term coined by
J.~Ottesen \cite{Ottesen}), in technical terms of a {\em transducer,
sensor, actuator} that becomes useful as soon as we understand the
underlying mathematical equations.

\subsection{Simulation and Prediction}\label{ss:simulation}
Once a model is found and verified and the system's parameters are
estimated for one domain, one has the hope of doing computer
``experiments" (i.e., calculations and extrapolations for modified
data input) to replace or supplement costly, time-consuming and
sometimes even physically impossible experiments. The last happens
when we are permitted to change a single parameter or a selected
combination of parameters in the calculation contrary to a real
experiment, where typically one change induces many accompanying
changes. In this way we may predict what we should see in  new
experiments in new domains (new materials, new temperatures etc).
Rightly, one has given that type of calculations a special name of
honor, {\em computer simulations}: As a rule, it requires to run the
process on a computer or a network of computers under quite
sophisticated conditions (discussed in \cite{Shillcock}). Typically,
the problem is to bring the small distances and time intervals of
well-understood molecular dynamics up to reasonable mesoscopic
scales, either by aggregation or by Monte Carlo methods -- as
demonstrated by Buffon's needle casting for the numerical
approximation of $\pi$.

One should be aware that the word ``simulation" has, for good and
bad, a connotation derived from NASA space simulators and Nintendo
war games and juke boxes. Animations and other advanced computer
simulations can display an impressive beauty and convincing power.
That beauty, however, is often their dark side: Simulations can show
a deceptive similarity with true observations, so for the lipid
bilayer fusion of an insulin vesicle with the plasma membrane and
the release of the bulk of hormone molecules. The numerical solution
of huge systems of Newton's equation, i.e., the integration of all
the forces between the membrane lipids can be tuned to display a
convincing picture of the secretion course in a nanosecond time span
whereas that very process in reality takes seconds and minutes. In
numerical simulation, like in mathematical statistics, results which
fit our expectations too nicely, must awake our vigilance instead of
being taken as confirmation.


\subsection{Control}\label{ss:control}
The prescriptive power of mathematization deserves a more critical
examination. The time will come when the model based understanding
of intracellular dynamics in healthy and dysfunctional $\beta$-cells
will lead to new diagnostic approaches, new drugs and new
treatments. In physics and engineering we may distinguish between
the (a) feasibility, the (b) efficiency, and the (c) safety of a
design. A {\em design} can be an object like an airplane or a
circuit diagram for a chip, an instrument like a digital
thermometer, TV set, GPS receiver or pacemaker, or a regulated
process like a feed-back regulation of the heat in a building, the
control of a power station, the precise steering of a radiation
canon in breast cancer therapy - or the design of a new,
non-symptomatic diagnostic procedure or therapy.

Mathematics has its firm footing for testing the {\em feasibility}
of new approaches in thought experiments, estimations of process
parameters, simulations and solving equations. For testing {\em
efficiency}, a huge inventory is available of mathematical quality
control and optimization procedures by variation of key parameters.
It seems to me, however, that safety questions provide the greatest
mathematical challenges. For early diagnosis, say of juvenile
diabetes (DT1) and drug design, mathematics does not enter trivially
into the certification of the correctness of the design copy and the
quality test of the performance. Neither do we come to a situation
where it suffices to modify and re-calculate well-established models
and procedures. Experienced pharmacologists and medical doctors, we
may hope, will not trust mathematical calculations and adaptations.
Too many parameters may be unknown and pop up later. Here is a
parallel to the early days of traditional railroad construction: A
small bridge was easily calculated and built, but then
photogrammetrically checked when removing the support constructions.
A lowering of more than $\delta_{\operatorname{crit}}$ required
re-building. Similarly, even the most carefully calculated and
clinically tested diagnoses and therapies will require supplement by
the most crazy mathematical imagination of what could go wrong and
might show up only after years of treatment and where and how to
find or build an emergency exit in the cell.

An additional disturbing aspect of science-integrated medical
technology development is the danger of losing transparency. Medical
doctors are trained to understand the elements of mechanics and
chemical reactions, i.e., purely locally in cell terms. They are not
prepared to grasp global cell phenomena like magnetic field density
and the geometry and dynamics of long-distance amplification
processes within cells. Therefore, it will be very unfortunate when
medical doctors shall ordinate a treatment they do not really
understand.

\subsection{Explain phenomena}
The noblest role of mathematical concepts in cell physiology is to
explain phenomena. Einstein did it in physics when {\em reducing}
the heat conduction to molecular diffusion, starting from the formal
analogy of Fick's Law with the cross section of Brownian motion. He
did it also when {\em generalizing} the Newtonian mechanics into the
special relativity of constant light velocity and again when {\em
unifying} forces and curvature in general relativity.

One may hope that new mathematical models can serve biomedicine by
reducing new phenomena to established physical principles; and as
heuristic devices for suitable generalizations and extensions.

Physics history has not always attributed the best credentials to
explaining phenomena by abstract constructions. It has discarded the
concept of a ghost for perfect explanation of midnight noise in old
castles; the concept of ether for explaining the finite light
velocity; the phlogiston for burning and reduction processes, the
Ptolemaic epicycles for planetary motion. It will be interesting to
see in the years to come whether some of the common explanations in
cell physiology will suffer
the same fate.

\subsection{Theory development}
Finally, what will be the role of mathematical concepts and
mathematical beauty for the very theory development in cell
physiology? Not every mathematical, theoretical and empirical
accumulation leads to theory development. Immediately after
discovering the high-speed rotation of the Earth around its own
axis, a spindle shape of the Earth was suggested and an
infinitesimal tapering towards the North pole confirmed in geodetic
measurements around Paris. Afterwards, careful control measurements
of the gravitation at the North Cap and at the Equator suggested the
opposite, namely an ellipsoid shape with flattened poles. Ingenious
mathematical mechanics provided a rigorous reason for that. Gauss
and his collaborator Listing, however, found something different in
their control. They called the shape {\em gleichsam
wellenf{\"o}rmig} and dropped the idea of a theoretically
satisfactory description. Since then we speak of a {\em Geoid}.

Similarly, when analyzing intracellular geometries and dynamics we
may meet events which have their own phylogenetic history, dating
back to more than 0.6 billion of years in the case  of
$\beta$-cells, and have lost their relevance since then. With high
probability, many of the phenomena we observe are heritage,
meaningless relics of past hundreds of millions of species'
development. Neglecting the probabilistic ruin character of our
existence and pressing it into a slick mathematical model may be
quite misleading.

\section{Non-Invasive Control of Magnetic Nanoparticles}

\subsection{Emerging Radically New Research Agenda}
Addressing the intracellular geometry and dynamics of the cell has
many levels and many scales. To give an example, I shall describe an
evolving - focussed - systems biology of regulated exocytosis in
pancreatic $\beta$-cells, mostly based on \cite{Beta}. These cells
are responsible for the appropriate insulin secretion. Insufficient
mass or function of these cells characterize Type 1 and Type 2 {\em
diabetes mellitus} (DT1, DT2). Similar secretion processes happen in
nerve cells. However, characteristic times for insulin secretion are
between 5 and 30 minutes, while the secretion of neurotransmitters
is in the millisecond range. Moreover, the length of a $\beta$-cell
is hardly exceeding 4000 nanometres (nm), while nerve cells have
characteristic lengths in the cm and meter range. So, processes in
$\beta$-cells are more easy to observe than processes in nerve
cells, but they are basically comparable.

It seems that comprehensive research on $\beta$-cell function and
mass has been seriously hampered for 80 years because of the high
efficiency of the symptomatic treatment of DT1 and DT2 by insulin
injection. Recent advances - and promises - of noninvasive control
of nanoparticles suggest the following radically new research
agenda, to be executed first on cell lines, then on cell tissue of
selected rodents, finally on living human cells:



\subsubsection{Optical Tracking of Forced Movement of Magnetic Nanoparticles}
Synthesize magneto-luminescent nanoparticles; develop
a precisely working electric device, which is able to generate a
properly behaving electromagnetic field; measure cytoskeletal
viscosity and detect the interaction with organelles and actin
filaments by optical tracking of the forced movement of the
nanoparticles. Difficulties to overcome: protect against protein
adsorption by suitable coating of the particles and determine the
field strength necessary to distinguish the forced movement from the
underlying Brownian motion.


\subsubsection{Optical Tracking of the Intracellular Dynamics of Insulin Granules}
Synthesize luminescent nanoparticles with after-glow
property (extended duration of luminescence and separation of
excitation and light emission); dope the nanoparticles with suitable
antigens and attach them to selected organelles to track
intra-cellular dynamics of the insulin granules.


\subsubsection{Precise Chronical Order of Relevant (Electrical) Secretion Events}
Apply a multipurpose sensor chip and measure all electric phenomena,
in particular varying potentials over the plasma membrane, the
bursts of $\operatorname{Ca}^{2+}$ ion oscillations,  and changing
impedances on the surface of the plasma membrane for precise
chronical order of relevant secretion events.

\subsubsection{Geometry and Dynamics of Lipid Bilayer Membrane-Granule Fusion}
Describe the details of the bilayer membrane-granule fusion event
(with the counter-intuitive inward dimple forming and hard numerical
problems of the meso scale, largely exceeding the well-functioning
scales of molecular dynamics).


\subsubsection{Connecting Dynamics and Geometry with Genetic Data}
Connect the preceding dynamic and geometric data with
reaction-diffusion data and, finally, with genetic data.

\subsubsection{Health Applications}
Develop clinical and pharmaceutical applications:
\begin{itemize}
\item Quality control of transplants for DT1 patients.
\item Testing drug components for $\beta$-cell repair.
\item Testing nanotoxicity and drug components for various cell types.
\item Early in-vivo diagnosis by enhanced gastroscopy.
\item Develop mild forms of gene therapy for patients with
over-expressed  major type 2-diabetes gene TCF7L2
by targeting short interfering RNA sequences (siRNAs) to the
$\beta$-cells, leading to degradation of excess mRNA transcript.
(This strategy may be difficult to implement, due to the degradation
of free RNA in the blood and the risk of off-target effects.)
\end{itemize}

In the following, we shall not deal with the envisaged true health
applications. Only shortly we shall comment upon the mathematical
challenges of the non-invasive control of magnetic nanoparticles,
the intricacies of the related transport equations, compartment
models, electromagnetic field equations, free boundary theory,
reaction-diffusion equations, data analysis etc.

\subsection{Gentle Insertion by Rolling on Cell Surface} \label{ss:rolling}
The good news is the newly developed {\em Dynamic Marker} technique,
see \cite{Koch}:  Based on well-established electrical power
engineering know-how, arrays of conventional coils are arranged in
small engines to generate precisely directed dynamic magnetic field
waves of low magnetic field density (mTesla range) and low frequency
(1-40 Hz). The use of dynamic and directed magnetic field waves
makes the beads roll on the cell surface to rapidly meet willing
receptors.

This technology has shown a much better insertion performance in
experiments than conventional diffusion or static magnetic fields:
With less than 10 minutes characteristic dynamic marking is much
faster than waiting for 12-24 hours on diffusion of the
nanoparticles across the plasma membrane and reduces dramatically
the inflammation risks during waiting. Contrary to the conventional
use of static magnetic fields (e.g., by applying MRI machines) for
transfection the cell's nucleus will not be ``bombed" and the
associated high lysis (= break down) of cells under the process is
avoided. Success has only been achieved, however, for magnetic
nanoparticles with a diameter $\le 100 \operatorname{nm}$. Moreover,
the success seems to depend heavily on the correct tuning of
magnetic field density and frequency. The guiding transport
equations seem not fully understood yet. To generate much higher
magnetic field densities of field waves for in-vivo application,
portable superconductive coils at relatively high temperature
($90\/^o$ K) are expected to be developed in the future.

\subsection{Viscosity in Newtonian and Non-Newtonian
Cytosol}\label{ss:viscosity} Like many other biomedical quantities,
the viscosity of the cytosol cannot be measured directly. Let us
look at the eight to twelve thousand densely packed insulin vesicles
in a single $\beta$-cell. They all must reach the plasma membrane
within a maximum of 30 minutes after stimulation, to pour out their
contents. Let us ignore the biochemistry of Figure \ref{f:motility1}
and the many processes taking place simultaneously in the cell and
consider only the basic physical parameter for transport in liquids,
namely the viscosity of the cell cytosol. From measurements of the
tissue (consisting of dead cells) we know the magnitude of viscosity
of the protoplasma, namely about 1 milli-pascal-seconds (mPa s),
i.e., it is of the same magnitude as water at room temperature. But
now we want to measure the viscosity in living cells: before and
after stimulation; deep in the cell's interior and near the plasma
membrane; for healthy and stressed cells.

It serves no purpose to kill the cells and then extract their
cytosol. We must carry out the investigation in vivo and in loco, by
living cells and preferably in the organ where they are located. The
medical question is clear. So is the appropriate technological
approach with noninvasive control of magnetic nanoparticles,
explained in Section \ref{ss:rolling} above. These particles are
primed with appropriate antigens and with a selected color protein,
so that their movements within the cell can be observed with a
confocal multi-beam laser microscope which can produce up to 40
frames per second. The periods of observations can become relatively
short, down to 8-10 minutes - before these particles are captured by
cell endosomes and delivered to the cells' lysosomes for destruction
and consumption of their color proteins (see also below Section
\ref{ss:apoptosis}).

\subsubsection{Newtonian Idealization} Assuming (wrongly, see
below) that the cytosol is a Newtonian liquid, we get from A.
Einstein \cite{Ein} and M. von Smoluchowski \cite{Smo} precise
recipes how to determine the viscosity from a few snapshots of the
Brownian motion or the forced movement of suspended particles.
Roughly speaking, Einstein discovered the scale independence
(self-similarity = fractal structure) of the Brownian motion
permitting to derive the characteristic diffusion coefficients of
the almost continuously happening jumps, from sample geometries of
positions at observable, realistic huge time intervals (huge
compared to the characteristic time of the process). Smoluchowski
worked out a smart observation scheme in the case of the presence of
many particles which can no longer be traced individually by then
(and now) existing equipment.

More precisely, the simplest mathematical method to determine its
viscosity {\em in vivo} would be just to pull the magnetized
particles with their fairly well-defined radius $a$ with constant
velocity $v$ through the liquid and measure the applied
electromagnetic force $F$. Then the viscosity $\eta$ is obtained
from Stokes' Law $F=6\pi a\eta v$. The force and the speed must be
small so as not to pull the particles out of the cell before the
speed is measured and kept constant. Collisions with insulin
vesicles and other organelles must be avoided. It can only be
realized with a low-frequency alternating field. But then Stokes'
Law must be rewritten for variable speed, and the mathematics begins
to be advanced. In addition, at low-velocity we must correct for the
spontaneous Brownian motion of particles. Everything can be done
mathematically: writing the associated stochastic Langevin equations
down and solve them analytically, or approximate the solutions by
Monte Carlo simulation, like in \cite{Lea, Schw}. However, we
rapidly approach the equipment limitations, both regarding the laser
microscope's resolution and the lowest achievable frequency of the
field generator.

So we might as well turn off the field generator and be content with
intermittently recording the pure Brownian motion $\bm{x}(t)\in
\mathbb{R}^3$ of a single nanoparticle in the cytosol! As shown in
the cited two famous 1905/06 papers by Einstein, the motion's
variance (the mean square displacement over a time interval of
length $\tau$) $\sigma^2=\langle \bm{x}^2\rangle =
E(|\bm{x}(t_0+\tau)- \bm{x}(t_0)|^2)$ of a particle dissolved in a
liquid of viscosity $\eta$ is given by $\sigma^2=2D\tau$, where $D =
k_BT/(6\pi a\eta)$ denotes the diffusion coefficient with Boltzmann
constant $k_B$\/, absolute temperature $T$ and particle radius $a$.
In statistical mechanics, one expects $10^{20}$ collisions per
second between a single colloid of $1 \mu m$ diameter and the
molecules of a liquid. For nanoparticles with a diameter of perhaps
only 30 nm, we may expect only about $10^{17}$ collisions per
second, still a figure large enough to preclude registration. There
is simply no physical observable quantity $\langle \bm{x}^2\rangle$
at the time scale $\tau = 10^{-17}$ seconds. But since the Brownian
motion is a Wiener process with self-similarity we get approximately
the same diffusion coefficient and viscosity estimate, if we, e.g.,
simply register 40 positions per second. Few measurements per second
are enough. Enough is enough, we can explain to the experimentalist,
if he/she constantly demands better and more expensive apparatus.

Note that $\sigma^2$  also can be estimated by the corresponding
two-dimensional Wiener process of variance $3/2\,\sigma^2$\/,
consisting of the 2-dimensional projections of the 3-dimensional
orbits, as the experimental equipment also will do.

Now you can hardly bring just a single nanoparticle into a cell.
There will always be many simultaneously. Thus it may be difficult
or impossible to follow a single particle's zigzag path in a cloud
of particles by intermittent observation. Also here, rigorous
mathematical considerations may help, namely the estimation of the
viscosity by a periodic counting of all particles in a specified
{\em window}. As mentioned above, the necessary statistics was done
already in \cite{Smo}.

\subsubsection{Non-Newtonian Reality} Beautiful, but it is still
insufficient for laboratory use: There we also must take into
account the non-Newtonian character of the cytosol of $\beta$-cells.
These cells are, as mentioned, densely packed with insulin vesicles
and various organelles and structures. Since the electric charge of
iron oxide particles is neutral, we can as a first approximation
assume a purely elastic impact between particles and obstacles. It
does not change the variance in special cases, as figured out for
strong rejection of particles by reflection at an infinite plane
wall in \cite{Smo}. But how to incorporate the discrete geometry and
the guiding role of the microtubules into our equations of motion?
Here also computer simulations have their place to explore the
impact of different repulsion and attraction mechanisms on the
variance.

\subsection{Sensing Microfilaments, Tracing Insulin Granule Motility,
Displaying the Genetic Variety of the Diabetes Umbrella} Up to now,
it is not clear what geometry or geometries are underlying the
secretion dynamics.

\subsubsection{Pressing Need for Geometric Invariants}
Soon we may be able to sense and map the extended
geometry of the microtubules and their smoothing before secretion;
soon we may be able to sense and map the extended geometry of the
actin filaments and their dissolution just before secretion.
However, we shall need numbers or other mathematical objects to
characterize observed geometries and dynamics in order to relate
observations of well-functioning secretion and dysfunction to the
effect of selected genes. Epidemiological studies in large
populations (see \cite{Lyssenko-Groop, Pal-Gloyn}) have found more
than 20 gene deviations which show up in families with high
expression of DT1 or DT2. Form that we learned that DT1 and DT2 are
not two single diseases (distinguished by simple symptomatic
classification) but umbrellas of quite different dysfunctions
leading to the same or similar symptoms. To give these patients a
cure, we need to know more precisely what is going wrong on the cell
level. Proteomic analysis, in particular the disclosure of the
proteins a certain gene is coding for, is a very promising approach.
However, it will only lead to success if we become able to
supplement it by precise description of related deviations in
geometry and dynamics.

\begin{figure}
\includegraphics[scale=0.6]{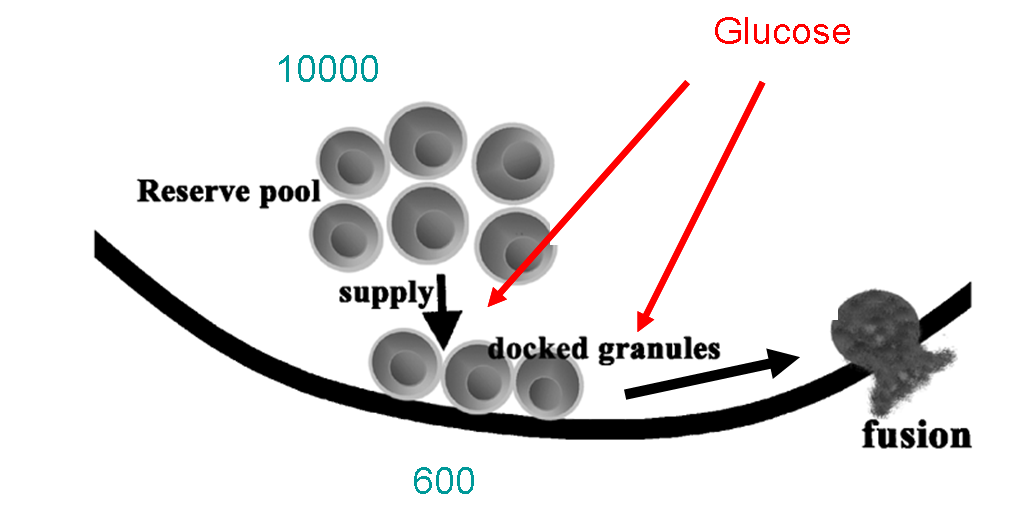}

\vspace{0.8cm}

\includegraphics[scale=0.68]{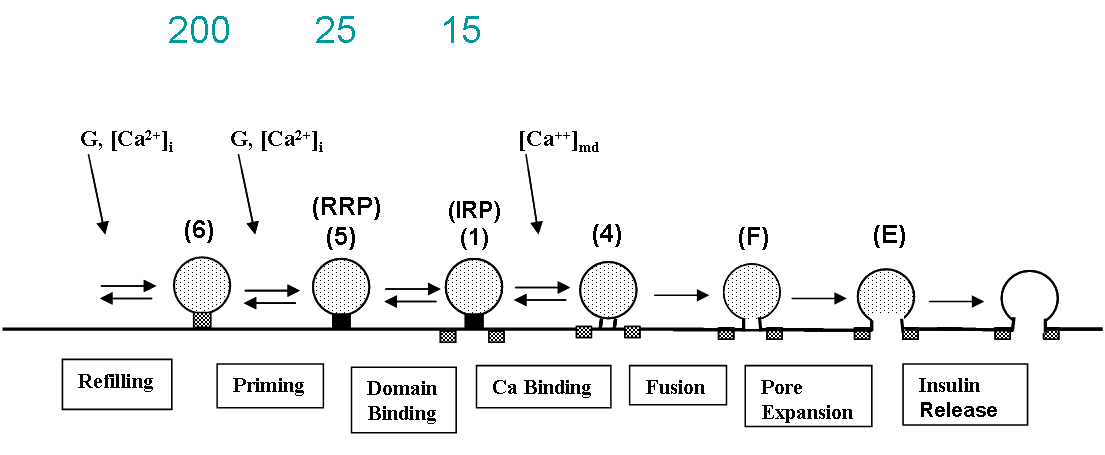}
\caption{a) Up: Basic two-pool model for exocytosis of Grodsky.
\newline b) Down: Extended six-pool model, incorporating Ca-binding,
of Chen, Wang and Sherman} \label{f:sherman}
\end{figure}

\subsubsection{First Steps via Compartment Models and Transition
Rates} Just as in engineering, economics or anywhere else, also in
cell physiology the daily mathematical exercise consists of the
estimation of some parameters; testing the significance of some
hypotheses; and designing compartment models for the dynamics of
coupled quantitative variables. A first step towards integrating
spatial geometry and temporal dynamics are the {\em Compartment
Models} of regulated exocytosis, first introduced by Grodsky
\cite{Gro} in 1972. He assumed that there are two compartments
(pools) of insulin granules, docked granules ready for secretion and
reserve granules, see Figure \ref{f:sherman}a. By assuming suitable
flow rates for outflow from the docked pool and resupply from the
reserve pool to the docked pool, the established biphasic secretion
process of healthy $\beta$-cells (depicted in Figure
\ref{f:biphasic}) could be modeled {\em qualitatively} correct. By
extending the number of pools from two to an array of six (Figure
\ref{f:sherman}\/b) and properly calibrating all flow rates, Chen,
Wang and Sherman in \cite{CWS} obtained a striking {\em
quantitative} coincidence with the observed biphasic process, see
also Toffolo, Pedersen, and Cobelli \cite{ToPeCo}. Such compartment
models invite the experimentalists (both in imaging and in
proteomics) and the theoretician (both in geometry and in
mathematical physics) to verify the distinction of all the
hypothetical compartments in cell reality and to assign global
geometrical and biophysical values to the until now only tuned flow
rates. A geometer who is familiar with mathematical physics may, for
instance, look for small inhomogeneities in the cell which could
drive the global dynamics. One candidate is the viscosity as
addressed above in Section \ref{ss:viscosity}: A slight increase of
cytosol viscosity from the cell interior towards the plasma membrane
would induce an externally directed vector by diffusion statistics.

A self-imposed limitation is the low spatial resolution of the
aggregated compartments, which does not allow to investigate the
local geometry and the energy balance of the secretion process.

\subsection{Electrodynamic Insulin Secretion ``Pacemaker"}
In the preceding sections I briefly described the common
phenomenological approaches to regulated exocytosis: the focus on
the variable discrete geometry of the microfilaments; the numerical
treatment of the molecular dynamics visualizing the singularity of
lipid bilayer fusion events; the analytic power of compartment
models to reproduce biphasic secretion.  The phenomenological
approaches relate the various data by visible evidence and
statistically more or less well supported ad–hoc assumptions about
the regulation. They focus on the dominant and visible structures
(like the filaments) and measurable local states in the neighborhood
of the fusion event, neglecting long-distance phenomena like
electromagnetic waves across the cell. One may deplore that and also
the ``enormous gap between the sophistication of the models and the
success of the numerical approaches used in practice and, on the
other hand, the state of the art of their rigorous understanding"
(Le Bris \cite{LebICM} in his 2006 report to the International
Congress of Mathematicians).

In \cite{Apu}, the authors advocate for supplementing the
phenomenological approach by a theoretical approach based on first
principles, following Y. Manin's famous parole: {\em The visible
must be explained in terms of the invisible}, \cite{Manin}. They
explain how a combination of rigorous geometrical and stochastic
methods and electro-dynamical theory naturally draws the attention
to fault-tolerant signalling, self-regulation and {\em
amplification} (in M. Gromov's terminology). They consider the
making of the fusion pore, anteceding the lipid bilayer membrane
vesicle fusion of regulated exocytosis, as a {\em free boundary
problem} and show that one of the applied forces is generated by
glucose stimulated intra-cellular $\operatorname{Ca}^{2+}$ ions
oscillations (discussed in \cite{FrPh}) resulting in a low-frequent
electromagnetic field wave. It mis suggested that the field wave is
effectively closed via the weekly magnetic plasma membrane
(containing iron in the channel enzymes). Unfortunately, textbook
electrodynamics mostly deals with geometrically simple
configurations where the beauty and strength of Maxwell Equations
best come out, but is less informative for treating peculiar
geometries. This was also noted by Gelfand:
\begin{quotation}
... images play an increasingly important role in modern life, and
so geometry should play a bigger role in mathematics and in
education. In physics this means that we should go back to the
geometrical intuition of Faraday (based on an adequate geometrical
language) rather than to the calculus used by Maxwell. People were
impressed by Maxwell because he used calculus, the most advanced
language of his time. (\cite[p.\/xx]{Gelfand})
\end{quotation}

The recent experimental evidence of the bio-compatibility and
bio-efficiency of low-frequent electromagnetic fields (addressed
above in Section \ref{ss:rolling}) gives a hint to the presence of
global fields controlling local events in the cell and supports a
vision of a possible future electrodynamic pacemaker to stimulate
regulated exocytosis in tired, dysfunctional $\beta$-cells.

\subsection{Induced Apoptosis Chain Reaction in Cancer
Cells}\label{ss:apoptosis} The continuing almost total lack of
understanding of the global aspects of cell physiology can also lead
to happy surprises: In the course of the insertion experiments
described above in Section \ref{ss:rolling}, it was discovered that
the inserted iron oxide nanoparticles of diameter $< 50$ nm (with
special antibody-conjugated surfaces) were immobilized in less than
10 minutes within the lysosomes (organelles for digestion and
destruction). However, continuing the action of the low frequency
electromagnetic oscillations tore the membranes of these lysosomes,
purely mechanically.

That was bad news for exploring the intracellular geometry and
dynamics by nanoparticle transducers, because the observation window
is consequently short, only 10 minutes if we come to use the
``wrong" antibodies. It was good news for cancer research:
Destroying the lysosomes in a probe of cancer cells leads to release
of the digestive enzymes and initiates a destructive chain reaction
in the neighboring tissue which stops automatically when healthy
tissue (with neutral pH) is reached. The range (in time and space)
of the obtainable chain reactions has not yet been fully determined.
Nanotoxicity for healthy tissue, however, will be excluded
conclusively. Testing of the field generator is under preparation
for curing skin cancer on model tissue, on model animals and for
full proof of concept.

\section{Conclusions}

Down  there, in the nano-world of regulated exocytosis in pancreatic
$\beta$-cells, a heap of geometrical and dynamical information is
waiting to be interrelated. Encouragement and, perhaps, inspiration
may be gained from the visionary \cite{Carbone-Gromov} (though
restricted to molecular biology). Every mathematician's conviction
is the {\em inseparability of geometry and dynamics}. That's what we
teach the students in algebra classes with the concepts of {\em
orbits and ideals}; in ordinary differential equations classes with
the {\em Poincar{\'e}-Bendixson Theorem} and the significance of the
multiplicity and sign of eigenvalues for global behavior and the
geometry of bifurcations; and most outspoken in spectral geometry
classes with our focus on {\em spectral invariants} that
characterize both shape and change at the same time.

For the evolving medical biology of highly differentiated cells like
the pancreatic $\beta$-cells it remains to hope that tendencies to
futile overspecialization and excessive reductionism can be
overcome. Clearly, the base of all future advances must be the
precise, controlled single observations. But a real hope for
diabetes patients can only come from the integration of the already
established {\em local} facts into a {\em global} geometric and
dynamic perception.

\bibliographystyle{amsalpha}

\end{document}